\RequirePackage{fix-cm}
\documentclass[smallextended]{svjour3}       

\smartqed  
\usepackage{graphicx}
\usepackage{braket} 
\usepackage{physics} 
\usepackage{amsmath} 
\usepackage{mathtools} 
\graphicspath{ {images/} } 
\usepackage{color}

\begin{document}

\title{Relating quantum incoherence, entanglement and superluminal signalling.}



\author{Stanislav Filatov\and Marcis Auzinsh}

\authorrunning{Short form of author list} 

\institute{Stanislav Filatov \at
              Department of Physics,
 University of Latvia, Raina boulevard 19, 
 LV-1586, Riga, Latvia \\        
           \email{sfilatovs@gmail.com}            \\
          \and
           Marcis Auzinsh \at
              Department of Physics,
 University of Latvia, Raina boulevard 19, 
 LV-1586, Riga, Latvia \\        
           \email{Marcis.Auzins@lu.lv}   
}

\date{Received: date / Accepted: date}

\maketitle

\begin{abstract} 

Hereby we inspect two-partite entanglement using thought experiment that relates properties of incoherently mixed states to the impossibility of faster-than-light (FTL) signalling. We show that if there appears a way to distinguish ensembles of particles that are described by the same density matrix, but are generated using different pure states - properties of entanglement (namely, non-classical correlations) could be employed to create an FTL signalling device. We do not claim FTL signalling is possible, rather, we establish the logical connection between the aforementioned properties of current physical theory which has not so far been evident.

\end{abstract}
\section{Introduction}

The idea of the possibility of FTL (Faster-than-light) movement of some sort of information started to loom over quantum mechanics since the famous Einstein-Podolsky-Rosen paradox has been put forward \cite{EPR35}. Correlations between spacelike separated quantum particles are exceeding those possible in Classical Mechanics \cite{Bell64}. It seems like, if we assume that the measurement collapses the wavefunction of the particle, that measurement (and hence the choice of the basis of measurement) of one particle of the entangled pair immediately collapses the wavefunction of the other particle to the appropriate state. It seems like particles could somehow ``communicate", although being spacelike separated. Some researchers have even tried to put the lowest limit on the speed of such communication and got figures of about $10^7c$ \cite{Tittel98}. Before we continue, we should notice that separation of the entangled pair of particles into {\em first particle} and {\em second particle} is a bit imprecise jargon. Strictly speaking particles in an entangled pair are inseparable and we are not measuring first particle and saying that it influences the other. Rather we are making measurement on the whole entangled state of two particles whenever any detector detects one particle of an entangled state.

Still, naive attempts to create FTL communication device fail because after more careful investigation one can realize that the correlations between measurements are conditional and one sees the correlations only when the results of the measurements on both ends of the entanglement are compared. Hence, the result of one of the measurements has to be communicated through the classical channel whose speed is capped at the speed of light, this communication speed in a classical channel determines the overall speed of exchange of information. Measurement on only one side results in absolutely random outcomes described by the Identity Density matrix. See appendix A for calculation of density matrix for the measurement of one of the particles of the singlet state   \cite{Susskind}.

This paradox of non-locality has been the trigger for a large body of research in different directions. One of them is a two-sided process in which some researchers invent potential FTL communication devices \cite{Petroni80,Selleri79,Herbert79,Popper82,Herbert82,Greenberger98,Shiekh06} and others explain why those devices are unfeasible \cite{Ghirardi79,Ghirardi07,Ghirardi12,Ghirardi13,Wigner52,Araki61,Yanase61,Ghirardi81,Ghirardi82}. This process has not yet led to the development of actual FTL communication device, but has resulted in a number of important theorems for Quantum Information, one of the most famous being the No-cloning theorem \cite{Wotters82,Dieks82}. One can look at the references cited above for a detailed discussion, but in short, usually those are implicit assumptions that are incompatible with quantum mechanics that make the proposed FTL communication devices dysfunctional. Some do not take into account all the subtleties of the conservation laws of the composite system of the measured object and apparatus \cite{Petroni80,Selleri79,Herbert79}. Others create models which imply non-unitary transformation (existence of an operator that maps two distinct initial states onto the same final state)\cite{Shiekh06,Greenberger98}. And some assume a non-linear quantum process \cite{Greenberger98} which is also in contradiction with quantum mechanics as we know it now. 

Another direction for research are more general FTL impossibility proofs \cite{Ghirardi80,Ghirardi88,Gisin01}, which, in turn allow to put stricter criteria on proposals of modification of quantum mechanics. One of the important results has been that although usually non-linear additions to Quantum Mechanics  result in the possibility of FTL signalling \cite{Gisin90,Weinberg89,Weinberg89x2,Gisin89}, there are ways to introduce some non-linear dynamics in Quantum Mechanics without the need to violate no-signalling condition \cite{Ghirardi86,Pearle90}. It has also been shown that it is possible to introduce preferred Frame of reference in which the information about entanglement would move faster-than-light without getting into paradox of retrocausality \cite{Ryff19}.
Moreover, implications of some specific ways of introducing preferred frame of reference for FTL quantum information transfer between entangled particles have also been investigated \cite{Scarani00,Suarez97,Gisin02}

Generally, although FTL communication using entanglement is still an impossibility, analysis of this  apparent contradiction between quantum mechanics and relativity has been a way to deepen the understanding of entanglement and its relativistic properties. 

In this work we want to emphasize the following point. If there appears a way to distinguish maximally mixed ensembles of particles that are described by the same Identity density matrix, but are generated using different pure states - then properties of entanglement could be employed to create an FTL signalling device. In other words, the fact that complete information about the quantum state is revealed only through repeated measurements  on an ensemble of particles is directly linked to the impossibility of FTL signalling (see Section \ref{concl} for a detailed discussion). 

The rest of the paper is structured as follows. In Section \ref{mms} we discuss maximally mixed states and show how the same mixed quantum state could be generated using different pure states. In Section \ref{device} we imagine a hypothetical device that could distinguish maximally mixed ensembles generated using different pure states. After that in Section \ref{assumption} we state our implicit assumption about the way the entangled state collapses when the state of one of the particles gets measured. The assumption does not contradict the generally accepted description of the process, only clarifies the details of the second particle collapse. In Section \ref{ftlsignalling} we put everything together and show how distinguishability of differently generated maximally mixed states leads to FTL signalling through the properties of entanglement. In Section \ref{summary} we summarize all the steps we have taken to make our point. And in Section \ref{concl} we reflect on our findings.




\section{Maximally mixed states} \label{mms}

\begin{figure}
    \centering
    \includegraphics[width=0.3\textwidth]{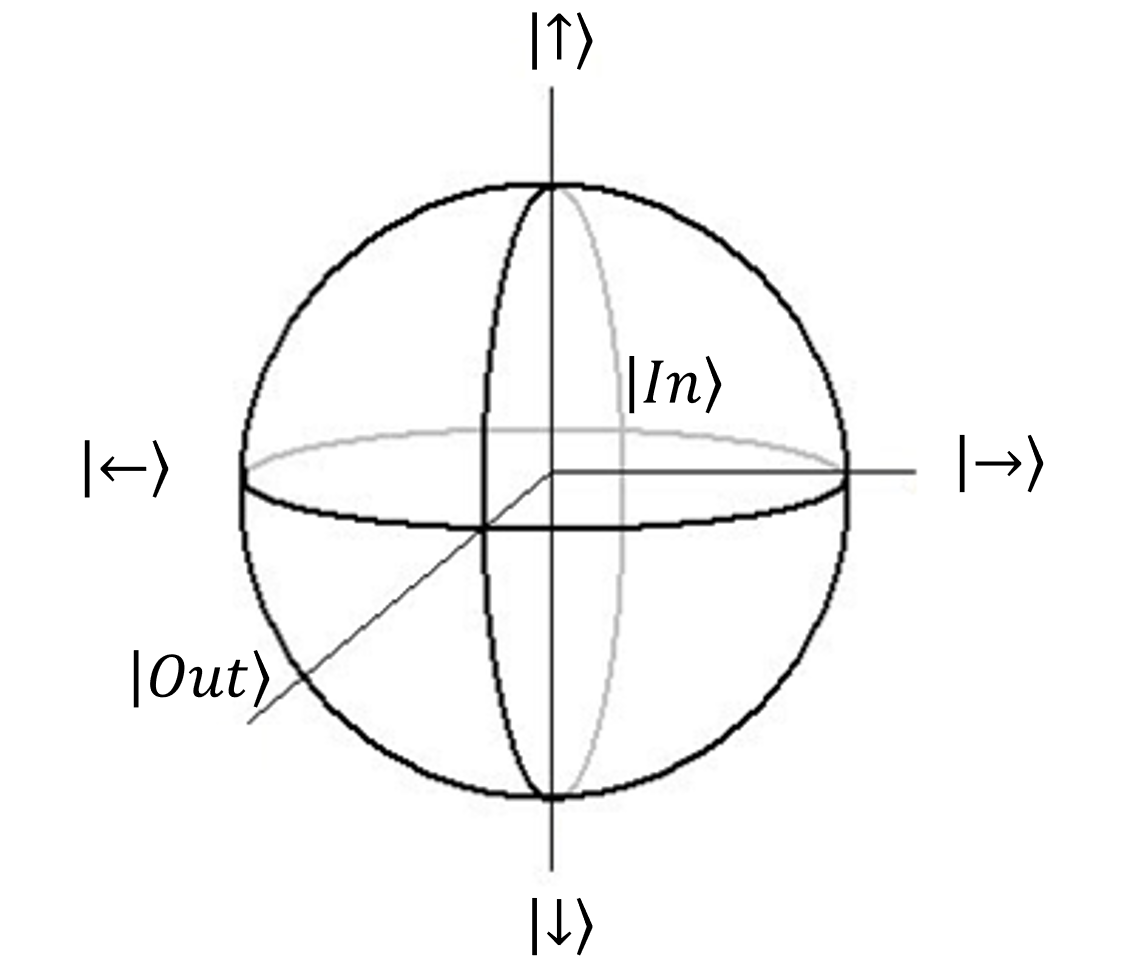}
    \caption{Bloch sphere representing a general two-level system. See eq. \ref{eq:transform} for formal relations between depicted states.}
    \label{fig:bloch-sphere}
\end{figure}

Let us consider a general quantum particle with two degrees of freedom (qubit). Throughout this work we will be using Bloch Sphere representation of a qubit (see fig. \ref{fig:bloch-sphere}). Physically there are many systems that could be described in this way, for example photonic polarization in $x$-$y$ plane, spin of an electron or even spatial superposition of a particle taking different paths inside an interferometer. Logic of this work applies to any two-level quantum system, so we will stick to the most straightforward notation, namely the one of the spin of an electron. $\ket{\uparrow}$ and $\ket{\downarrow}$ are states of the spin aligned along and opposite the quantization axis. All other states can be represented as superpositions of the abovementioned Hilbert-space orthogonal states, for example

\begin{equation}
\begin{split}
\ket{\rightarrow} = \frac{1}{\sqrt{2}}(\ket{\uparrow} +i \ket{\downarrow}) \\
\ket{\leftarrow} = \frac{1}{\sqrt{2}}(\ket{\uparrow} -i \ket{\downarrow}) \\
\ket{In} = \frac{1}{\sqrt{2}}(\ket{\uparrow} + \ket{\downarrow}) \\
\ket{Out} = \frac{1}{\sqrt{2}}(\ket{\uparrow} - \ket{\downarrow}) \\
\end{split}
\label{eq:transform}
\end{equation}

Pure states are located on the surface of the Bloch Sphere (we will represent them with arrows). Mixed partially coherent states are represented by a shorter arrow (that does not reach the surface) and length of the arrow is proportional to the amount of coherence of the state. Note that Hilbert-space orthogonal states (like $\ket{\uparrow}$ and $\ket{\downarrow}$) are pointing in opposite directions when represented on the Bloch sphere. Dot in the middle of the sphere is related to maximally mixed state. As we will be using this state extensively in our work it is worth discussing it in more detail.

Maximally mixed state, corresponding to the dot in the middle of the Bloch Sphere, corresponds (up to a normalization constant) to the Identity density matrix where off-diagonal terms (representing coherences) are equal to zero. We will be calling this state “I-rho state” (for “Identity-density-matrix state”). Usual interpretation attributes such a state to an ensemble of particles, not to each single particle. We will introduce it also in the same way \cite{GreenBook}. In order to generate an I-rho state we can take a device that spits out particles randomly with equal probability in all possible pure states. That would mean that Bloch Sphere representation of the particles coming out of this device is an arrow (reaching the surface because it is a pure state) that points in a random direction for each particle. Receiver of such particles, irrespectively of the basis he is measuring those, will see absolutely random results. Namely, if he chose any basis and measure the particles, he would get 50\% in first and 50\% in the second, orthogonal to the first, state.

\begin{figure}
    \centering
    \includegraphics[width=0.6\textwidth]{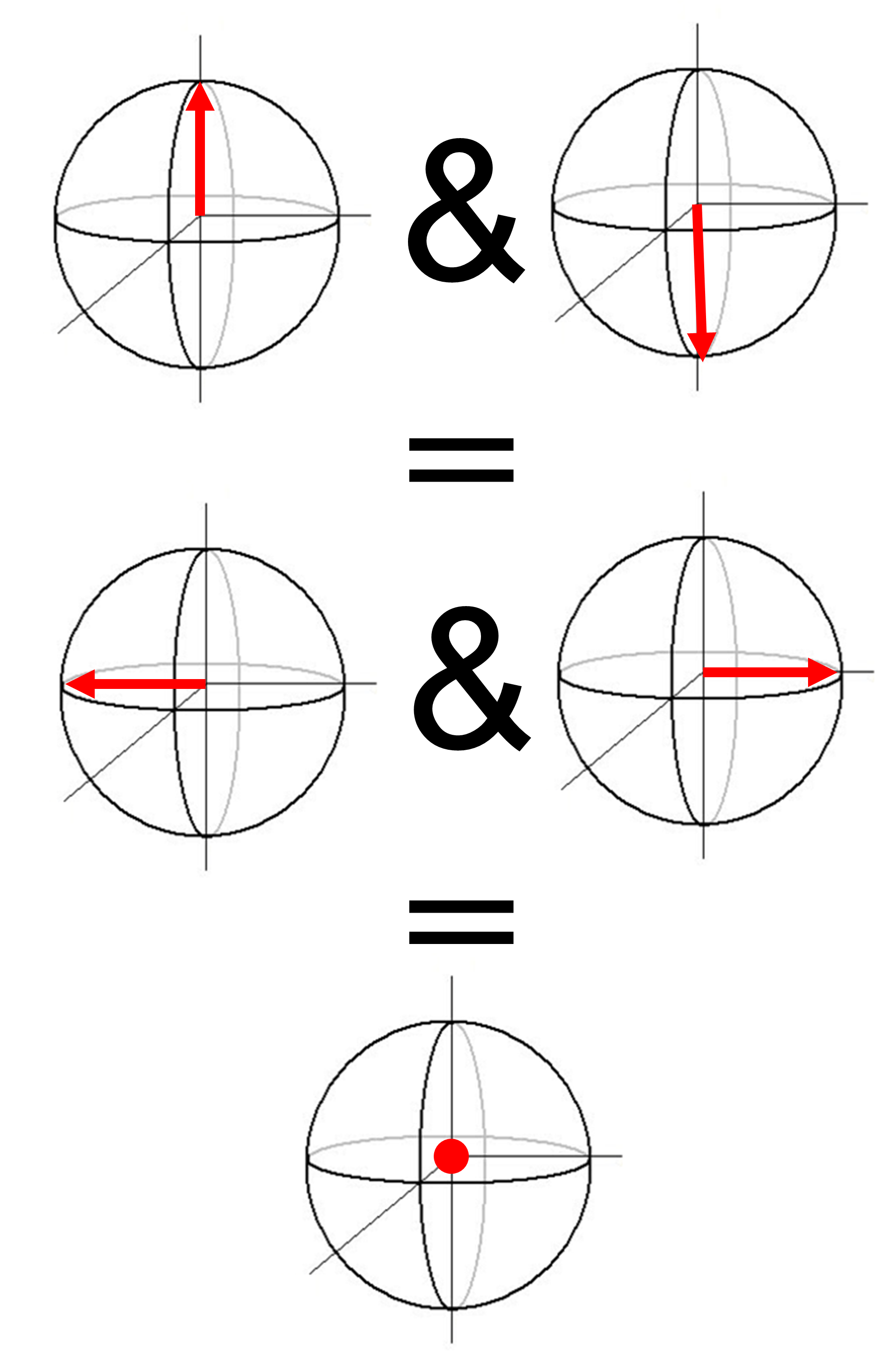}
    \caption{Bloch sphere representation of indistinguishability of I-rho states generated from different pure states. ``\&" sign in this case represents INCOHERENT sum, as described in our example. First line shows the case represented in eq. \ref{eq:rho1}, second line - the case from eq. \ref{eq:rho2}. Third line is the resulting state of Identity Density matrix.}
    \label{fig:irho-and}
\end{figure}

There are,  however, other ways to generate the I-rho state. Formally the states generated by this method are identical to the abovementioned (see eq. \ref{eq:rho1} and \ref{eq:rho2} for formal and fig. \ref{fig:irho-and} for visual description).  We can have a device that spits out randomly particles either (50\%) in state $\ket{\uparrow}$ or (50\%) in state $\ket{\downarrow}$. Another way could be to produce particles either (50\%) in state $\ket{\rightarrow}$ or (50\%) in state $\ket{\leftarrow}$. Generally, as long as the mixed state is generated using equal proportions of any two orthogonal states, they will result in the I-rho state. Formally all the three described ways of generation of I-rho state lead to the same density matrix for the ensemble. Still though, the pure states which are used in the two latter methods are different. We want to show what happens if the latter two methods of generation become distinguishable.

\begin{equation}
\begin{split}
    \rho_1 = &\frac{1}{2}(\ket{\uparrow}\bra{\uparrow} + \ket{\downarrow}\bra{\downarrow}) = \\
   &\frac{1}{2}\left(
    \begin{bmatrix} 1&\ 0\\ 0&\ 0 \end{bmatrix}
   + 
    \begin{bmatrix} 0&\ 0\\ 0&\ 1 \end{bmatrix}
    \right)
    = \\
    &\frac{1}{2}
    \begin{bmatrix} 1&\ 0\\ 0&\ 1 \end{bmatrix}
\end{split}
\label{eq:rho1}
\end{equation}

\begin{equation}
\begin{split}
    \rho_2 = &\frac{1}{2}(\ket{\rightarrow}\bra{\rightarrow} + \ket{\leftarrow}\bra{\leftarrow}) = \\
   &\frac{1}{4}\left(
    \begin{bmatrix*}[r] 1 & -i\\ i & 1 \end{bmatrix*}
   + 
    \begin{bmatrix*}[r] 1 &\  i\\ -i &\  1 \end{bmatrix*}
    \right)
    = \\
    &\frac{1}{2}
    \begin{bmatrix} 1&\ 0\\ 0&\ 1 \end{bmatrix}
    =
    \rho_1
\end{split}
\label{eq:rho2}
\end{equation}

\section{Device} \label{device}

Let us imagine for a moment a device that is able to make the following operation. It utilizes the fact that each method uses different conjugate basis in which to prepare the pure states of the particles. So it does not distinguish the pure states themselves, rather it distinguishes the conjugate basis in which the ensemble has been prepared. More precisely, given two ensembles of particles, generated in the same or in different ways, corresponding to the same I-rho state, it can tell whether the basis of the generation of the second ensemble is different from the first one.

Let’s say Bob has the Distinguishing device. Alice prepares first enseble of particles (N particles) in I-rho state in the ${\uparrow}$/${\downarrow}$ basis: randomly 50\% $\ket{\uparrow}$, 50\% ${\ket\downarrow}$ and sends them one-by-one to Bob, who collects all the particles into his device. Then Alice changes the basis of preparation of the I-rho state and randomly prepares next N particles either (50\% probability) in ${\ket\rightarrow}$  state or (50\%) in ${\ket\leftarrow}$. Sends them one-by-one to Bob who collects the particles into his device again. Device has a screen which prints 0 or 1 dependent on whether the basis of preparation of the second N-particle ensemble has been changed or not. In case the basis is changed the device would print 1. If Alice sent the second N-particle ensemble prepared in the same basis as the first one, the device would print 0. 

Although this device is not a physical possibility, it is a convenient way to introduce notion of distinguishability of maximally mixed ensembles generated using different pure states.

\section{Assumption about entanglement collapse after a measurement on a single particle} \label{assumption}

Before we show how properties of entanglement can be utilised to signal faster than light if there is a possibility to distinguish the bases of generation of the I-rho state, let us state the assumption we are making about the entangled particle state collapse. The assumption does not contradict the standard description of entangled state collapse, in fact, in the Introduction of \cite{Ghirardi13} similar reasoning is elegantly carried out in more formal and abstract way. Let us take the state 
\begin{equation}
{\frac{1}{\sqrt{2}}(\ket{\uparrow}\ket{\downarrow}-\ket{\downarrow}\ket{\uparrow})}
\label{eq:UD-DU}
\end{equation}
If we look at the state of a single particle, it is I-rho state which has equal probability of being any state (any point on a surface of a Bloch sphere), refer to Appendix for formal derivation. 

If we measure the entangled state and detect the first particle in particular eigenbasis, we make certain not only its state (the state of this particle is now depicted by an arrow reaching the surface of the Bloch Sphere), but also the state of the other particle (the other will definitely behave as if its state is an arrow reaching the bloch sphere in the opposite direction than that of the first one). This interpretation is consistent with the fact that if we measure both entangled particles in the same eigenbasis we get 100\% negative correlation and if the eigenbases are different, we get higher than classical correlations as predicted by quantum theory. If the second particle is measured by someone else who does not know about the measurement performed on the first particle, the fact that the entangled state wavefunction has been collapsed is impossible for him to notice because exact state into which the first particle collapses is absolutely random, so the state into which the second particle collapses is random as well. Hence, the measuring apparatus is still facing the same I-rho state. This I-rho state, however, is now generated using only pure states of a given eigenbasis (the eigenbasis of the measurement of the first particle). 

\begin{figure}
    \centering
    \includegraphics[width=0.9\textwidth]{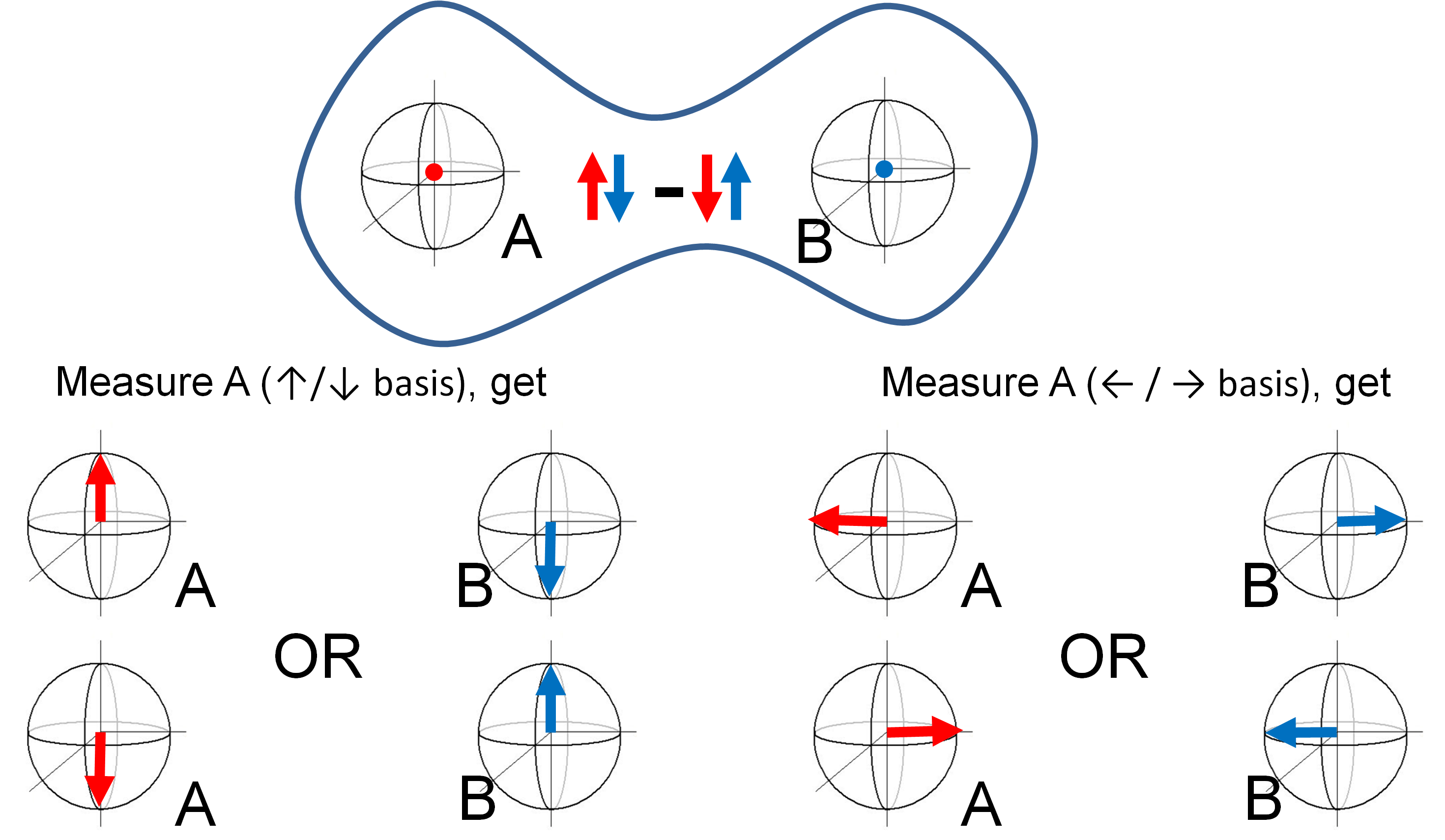}
    \caption{Bloch sphere representation of the process of measurement of the state of one of the particles of the entangled pair. A gets measured before B. Irrespective of the basis in which we measure A, B is in I-rho state. However, the I-rho state of B is composed of equal proportions of state $\ket{\uparrow}$ and ${\ket\downarrow}$ if A is measured in ${\uparrow}$/${\downarrow}$ basis, and the I-rho state of B is composed of equal proportions of state ${\ket\rightarrow}$  and  ${\ket\leftarrow}$ if A is measured in ${\rightarrow}$/${\leftarrow}$ basis}
    \label{fig:ent-meas}
\end{figure}

In short, (see Fig. \ref{fig:ent-meas}), when we collapse the first particle in a given eigenbasis, regardless of the particular eigenstate it appears in, the other particle of the entangled pair IMMEDIATELY behaves as if it has also been collapsed in the given eigenbasis (although it appears in random but opposite to the first eigenstate of that eigenbasis).

Therefore if we collapse the first particle in the ${\uparrow}$/${\downarrow}$ eigenbasis, its bloch sphere representation becomes either an ${\uparrow}$ arrow or ${\downarrow}$ arrow, and the representation of the second particle immediately becomes also either ${\uparrow}$ or ${\downarrow}$ arrow (facing the opposite direction). And  if we collapse the first particle in the ${\rightarrow}$/${\leftarrow}$ eigenbasis, its bloch sphere representation becomes either a ${\rightarrow}$  arrow or  ${\leftarrow}$ arrow, and the representation of the second particle immediately becomes also either ${\rightarrow}$  or  ${\leftarrow}$ arrow (facing the opposite direction). Although the states of the second particle in both cases correspond to the same I-rho state, the state itself is generated using different pure states.

\section{FTL signalling} \label{ftlsignalling}

Now let Charlie be the source of entangled particles in the singlet state who sends them to Alice and Bob. A and B are far away from each other and A operates on her particle before B is able to operate on his. There are ways to make sure detection of A happens earlier than B in all Lorentz frames by introducing a detour for the particle B as done in \cite{Ryff14}.

Alice holds projection device (polarizing beamsplitter with two detectors for example) that can be freely rotated to choose the reference basis so that she can collapse her particle in the eigenbasis of her choice. Bob holds the hypothetical I-rho - eigenbasis of generation - distinguishing device described in Section \ref{device}.

First N particles Alice collapses in ${\uparrow}$/${\downarrow}$ basis, next N in ${\rightarrow}$/${\leftarrow}$ basis. Bob’s particles immediately behave as if they have been also collapsed in the given eigenbasis (described in Section \ref{assumption}), so his device will be able to distinguish the two ensembles of N particles and would print “1”. As long as A and B are far away from each other, but the particles behave in the abovementioned way immediately after the collapse of one of the particles, FTL signalling becomes possible.

The N particles may even be sent one-by-one, as long as A and B are far away from each other and the time between first and Nth particle generation is less than the time it takes to send a signal from A to B.

\section{Summary} \label{summary}

Let us take a step back and see the whole logical path we have traversed. The exact state in which the particle has been prepared cannot be revealed by a measurement on a single particle; for that one needs many measurements performed on an ensemble of identically prepared particles. 

The other side of this property is that one can use ensemble to ``hide" the state of preparation of the particles. More precisely, one can prepare two ensembles of particles where the state of every single particle in ensemble 1 will be different from the state of every single particle in ensemble 2 (rigorously, the overlap between the states will not be 100\%). Moreover, the ensembles can be prepared in two different conjugate bases. Still, by carefully choosing the proportion of different states in the two ensembles, the ensembles could be made absolutely indistinguishable and the receiver of the particles, irrespectively of the way he is measuring those will face an incoherent maximally mixed state described by Identity density matrix.

We show that assumption of indistinguishability of those two differently prepared ensembles, if violated leads to possibility of faster-than-light signalling using the properties of entanglement. Non-classical correlations of entangled particles imply that IMMEDIATELY after the collapse of the statevector on one end of the entanglement, the other particle behaves as if it somehow ``knows" the eigenbasis of collapse of the first particle. The observer on the second end of the entanglement is therefore always facing the same Maximally mixed state, but prepared in the eigenbasis of the measurement on the first particle. If the information about the eigenbasis of preparation of such state is revealed via some hypothetical device, FTL signalling becomes possible. This could be done by entangling sufficient amount of pairs of particles, sending them to two far away receivers. The receiver A on one side collapses first half of the particles in one eigenbasis and the second half - in another. The receiver B on the other side would detect the change of eigenbasis of collapse using the abovementioned device. 

\section{Conclusions and Discussion} \label{concl}

We believe the main contribution of this work is in revealing the minimal way in which quantum mechanical description of the world should be violated so that FTL signalling using the properties of entanglement becomes possible. The key is being able to distinguish maximally mixed ensembles of particles that are generated in different eigenbases. In other words, when one is able to extract some information on the eigenbasis of preparation of single particles inside a maximally mixed ensemble, even at a cost of not being able to measure the states of the single particles themselves, he or she or it will be able to send signals FTL. And although such distinguishability, along with FTL signalling, are not possibilities within the physical theory as we know it now, the logical connection between properties of maximally mixed states, entangled pair collapse and FTL signalling is exciting and might be a guide for other researchers.

If we were to formulate our result in the most general way it would be following. 

1. in order to harvest entanglement to communicate FTL one needs to be able to determine the eigenbasis in which the system has been prepared. 

2. Statistical quality of Quantum Mechanics (namely the fact that full info about the state of the system can only be obtained through measurements on multiple instances of that system) is what prevents revelation of the eigenbasis of preparation in case of entangled particles.

3. Indistinguishability of maximally mixed states that, although generated using different pure states, correspond to the same density matrix description ensures that information on the eigenbasis of preparation of pure states is lost in the process of gathering statistics about the system. 

Interestingly, No-cloning theorem, especially when viewed in its historic context together with proposed cloning devices, could be described by the same three statements with third point modified as follows:

3. The fact that single instance of the system cannot be cloned in order to generate statistics about the system is what ensures that information on the eigenbasis of preparation is irretrievable on the level of a single particle. 

In fact, Herbert's famous FLASH cloning device \cite{Herbert82} which stimulated derivation of No-cloning Theorem (mentioned in the beginning of this work) can be seen as single-particle eigenbasis-of-preparation distinguishing device.

When looked from this angle, our result and No-cloning theorem seem to be, respectively, ensemble-level and single-particle-level manifestations of the statement ``Statistical properties of Quantum Mechanics prevent one from communicating FTL using entanglement". 

Additionally, we would like to mention our previous work \cite{Our19} where impossibility to distinguish the eigenbasis of preparation of the maximally mixed state (and therefore impossibility of FTL signalling) is used as a restriction on what could possibly happen (or not happen) during the process of interaction-free measurement (IFM) when both the object and the probe are quantum. Result being necessity of creation of entanglement during any process of quantum-quantum non-interaction, although from the first sight non-interaction should leave the states of the involved quantum particles unchanged.

\appendix
\section{Explicit calculation of the density matrix of one of the particles of a singlet state.}

In this appendix we are recreating derivation done in \cite{Susskind}. Although throughout the text we have used arrow notation, this particular derivation is more demonstrative if letter indexes are used, so we will be using $u/d$ instead of $\uparrow$/$\downarrow$. Let us imagine a measurement done only on one (first) of the particles of an entangled state. Physically this could mean that particles are far away from each other and we are making a measurement of the first particle. 

Two-particle state where states of each individual particle are discreet could be written in general form as follows:

\begin{equation}
\begin{split}
\sum_{a,b} \Psi_{ab}\ket{ab}     
\end{split}
\label{eq:GeneralPsi}
\end{equation}

Where $a$ - possible values of the first particle, $b$ - possible values of the second particle, $\Psi_{ab}$ and $\ket{ab}$ are, respectively, the coefficient and the statevector corresponding to different values of $a$ and $b$. If $a$ and $b$ are taking only 2 values each, say $u$ and $d$ (where $u$ and $d$ denote the spin of the particle being aligned along or in the opposite direction of the quantization axis), then the state for such a system takes the form:

\begin{equation}
\begin{split}
\Psi_{uu}\ket{uu} + \Psi_{ud}\ket{ud} + \Psi_{du}\ket{du} + \Psi_{dd}\ket{dd}     
\end{split}
\label{eq:Psi_ab}
\end{equation}

For the singlet state $\frac{1}{\sqrt{2}}(\ket{ud}-\ket{du})$ coefficients are 

\begin{equation}
\begin{split}
\Psi_{uu} =&\quad \ 0 \\
\Psi_{ud} =&\quad \: \frac{1}{\sqrt{2}} \\
\Psi_{du} =& -\frac{1}{\sqrt{2}} \\
\Psi_{dd} =&\qquad  0 \\
\end{split}
\label{eq:Psi_singlet}
\end{equation}

Density matrix for a single particle with two orthogonal states is a $2\cross2$ matrix of the form \\
\begin{equation}
\begin{split}
\rho_{aa'} =& \begin{bmatrix} \rho_{uu}&\ \rho_{ud}\\ \rho_{du}&\ \rho_{dd}\end{bmatrix}
\end{split}
\label{eq:rho_a}
\end{equation}

Note that subscripts correspond to the values of $a$ and $a'$ (values for the first particle) and do not involve the values of $b$ (values of the second particle). $b$ enters into the computation as an index over which we sum over. Explicit calculation of each entry of the density matrix is given by the following formula:

\begin{equation}
\begin{split}
\rho_{aa'} = \sum_{b}\psi_{ab}\psi_{a^{'}b}^{*}
\end{split}
\label{eq:rho_sum}
\end{equation}

Inserting all the values for $a$ and $a'$, summing over $b$ in each case and conjugating where necessary we get the following entries:

\begin{equation}
\begin{split}
    \rho_{uu} =& \Psi_{uu}\Psi_{uu}^\ast + \Psi_{ud}\Psi_{ud}^\ast = \\
    =& 0\cdot0 + \frac{1}{\sqrt{2}}\cdot\frac{1}{\sqrt{2}} = \\
    =& \frac{1}{2} \\
    \rho_{ud} =& \Psi_{uu}\Psi_{du}^\ast + \Psi_{ud}\Psi_{dd}^\ast = \\
    =& 0\cdot(-\frac{1}{\sqrt{2}}) + \frac{1}{\sqrt{2}}\cdot0 = \\
    =& 0   \\
    \rho_{du} =& \Psi_{du}\Psi_{uu}^\ast + \Psi_{dd}\Psi_{ud}^\ast = \\
    =&(-\frac{1}{\sqrt{2}})\cdot0 + 0\cdot\frac{1}{\sqrt{2}} = \\
    =& 0 \\    
    \rho_{dd} =& \Psi_{du}\Psi_{du}^\ast + \Psi_{dd}\Psi_{dd}^\ast = \\
    =& (-\frac{1}{\sqrt{2}})\cdot(-\frac{1}{\sqrt{2}}) + 0\cdot0 = \\
    =& \frac{1}{2} \\   
\end{split}
\label{eq:I-rho-explicit}
\end{equation}

Then placing the entries we get the density matrix description of the first particle only
\begin{equation}
\begin{split}
    \rho_{aa'} =& \begin{bmatrix} \frac{1}{2}&\ 0\\ 0&\ \frac{1}{2}\end{bmatrix} = \frac{1}{2}*I
\end{split}
\label{eq:I-rho-end}
\end{equation}

Which is I-rho maximally mixed state.

\begin{acknowledgements}
We would like to express gratitude to Vyacheslavs Kashcheyevs for insightful discussions about I-rho state and Denis Bezrukov for discussions about No-cloning Theorem. Also, Theoretical Minimum lecture series by L. Susskind for the lucidly explained concepts. Support of our families has been invaluable.
\end{acknowledgements}



\end{document}